\documentclass[aps,prl,twocolumn,superscriptaddress,showpacs]{revtex4-1}

\usepackage{amsmath}
\usepackage{graphicx}
\usepackage{calc}
\usepackage{bm}

\usepackage[T2A]{fontenc}
\usepackage[cp1251]{inputenc}
\usepackage[english]{babel}

\begin{document}

\title{Current-induced control of the polarization state in a polar metal based heterostructure SnSe/WTe$_2$}

\author{N.N. Orlova}
\author{A.V.~Timonina}
\author{N.N.~Kolesnikov}
\author{ E.V.~Deviatov}

\affiliation{Institute of Solid State Physics of the Russian Academy of Sciences, Chernogolovka, Moscow District, 2 Academician Ossipyan str., 142432 Russia}

\date{\today}

\begin{abstract}
The concept of a polar metal proposes new approach of current-induced polarization control for ferroelectrics. We fabricate SnSe/WTe$_2$  heterostructure to  experimentally investigate charge transport between two ferroelectric van der Waals materials with different polarization directions. WTe$_2$ is a polar metal with out-of-plane ferroelectric polarization, while SnSe ferroelectric semiconductor is polarized in-plane, so one should expect complicated polarization structure at the SnSe/WTe$_2$ interface. We study $dI/dV(V)$ curves, which  demonstrate sharp symmetric  drop to zero $dI/dV$ differential conductance at some threshold bias  voltages $\pm V_{th}$, which are nearly symmetric in respect to the bias sign. While the gate electric field is too small to noticeably affect the carrier concentration, the positive and negative threshold positions are sensitive to the gate voltage. Also,  SnSe/WTe$_2$ heterostructure shows re-entrant transition to the low-conductive $dI/dV=0$ state for abrupt change of the bias voltage even below the threshold values. This behavior can not be observed for single SnSe or WTe$_2$ flakes, so we interpret it as  a result of the SnSe/WTe$_2$ interface coupling.  In this case, some threshold value of the   electric field at the SnSe/WTe$_2$ interface is  enough to  drive 90$^\circ$ change of the initial SnSe in-plane polarization in the overlap region. The polarization mismatch leads to the significant interface resistance  contribution, analogously to the scattering of the charge carriers on the domain walls.   Thus, we demonstrate  polarization state control  by electron transport through the SnSe/WTe$_2$ interface.

\end{abstract}

\maketitle

\section{Introduction}

Recently,  ferroelectric van der Waals materials attract significant interest both for  the fundamental physics and for promising  applications in quantum sensors, new memory devices, and for the ferroelectric field-effect transistors~\cite{ferrmem,feFET,synap,review}. For the fundamental research, some of these materials represent a novel concept of the intrinsic polar metal~\cite{PM,pm1,pm2,pm3,pm4}. The latter can be regarded as a ferroelectric in metals, it is characterized by intrinsic conduction and inversion symmetry breaking~\cite{PM}.  

Due to the finite conductance in polar metals, they propose new approaches to control ferroelectric polarization. For example,  it is possible to control   polarization by  charge current or, vice versa, to control charge transport by the ferroelectric polarization. Out-of-plane polarization can be affected or even switched  in well conducting semimetals MoTe$_2$ and  WTe$_2$  by piezoresponse force microscopy (PFM)~\cite{MoTe,WTe}. Also, the initial in-plane polarization can be managed in SnSe and  SnTe semiconductors using  scanning tunneling microscope (STM) technique~\cite{SnSe,SnTe}.  Manipulation of charge transport by ferroelectric polarization is important for ferroelectric field-effect transistors (FeFET), as it has been demonstrated for In$_2$Se$_3$-based structures~\cite{feFET}. Ferroelectric polarization can be controlled also by the in-plane current-induced electric field in WTe$_2$ and SnSe thin films~\cite{WTeour,SnSeour1,SnSeour2}. Van der Waals materials also offer strain control of ferroelectric polarization~\cite{MoTestrain}.

In the sense of the ferroelectric polarization control, it is quite natural to consider also van der Waals heterostructures with one or several ferroelectric materials and (possibly) some other layers. For example, MoS$_2$/h-BN/graphene/CuInP$_2$S$_6$  heterostructure has been proposed for long-retention memory~\cite{long-retent}, while GeSe/MoS$_2$ heterojunction represents novel FeFET realization~\cite{VHJ}. For the heterostructures with polar metals, theory predicts~\cite{PM} realization of multiple states with different relative directions of polar displacements due to the interface coupling effects. If both the ferroelectric materials are conducting, like for WTe$_2$ and SnSe, the interface coupling, and, therefore, the polarization state, could be  supposed to be controlled directly by electron transport through the heterostructure. 

Here, we fabricate SnSe/WTe$_2$  heterostructure to  experimentally investigate charge transport between two ferroelectric van der Waals materials with different polarization directions. WTe$_2$ is a polar metal with out-of-plane ferroelectric polarization, while SnSe ferroelectric semiconductor is polarized in-plane, so one should expect complicated polarization structure at the SnSe/WTe$_2$ interface. We study $dI/dV(V)$ curves, which  demonstrate sharp symmetric  drop to zero $dI/dV$ differential conductance at some threshold bias  voltages $\pm V_{th}$, which are nearly symmetric in respect to the bias sign. While the gate electric field is too small to noticeably affect the carrier concentration, the positive and negative threshold positions are sensitive to the gate voltage. Also,  SnSe/WTe$_2$ heterostructure shows re-entrant transition to the low-conductive $dI/dV=0$ state for abrupt change of the bias voltage even below the threshold values. This behavior can not be observed for single SnSe or WTe$_2$ flakes, so we interpret it as  a result of the SnSe/WTe$_2$ interface coupling. In this case, some threshold value of the   electric field at the SnSe/WTe$_2$ interface is  enough to  drive 90$^\circ$ change of the initial SnSe in-plane polarization in the overlap region. The polarization mismatch leads to the significant interface resistance  contribution, analogously to the scattering of the charge carriers on the domain walls.
 
\section{Samples and techniques}

\begin{figure}[t]
\center{\includegraphics[width=\columnwidth]{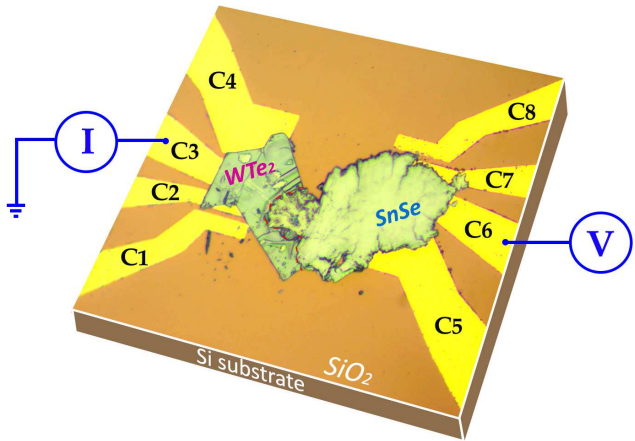}}
\caption{(Color online) Optical image of the WTe$_2$/SnSe sample with electrical connections. 100 nm thick Au leads pattern is formed on a top of the standard oxidized silicon wafer. Thin WTe$_2$ flake is placed above the left group of the Au leads (C1--C4), then SnSe flake is situated to overlap both the WTe$_2$ flake and the right leads (C5--C8). The Au leads outline can be seen under the flakes, so they are below 200~nm thickness. The right and the left leads are separated by 80~$\mu$m interval, the flakes are about the same lateral size. Red dashed line denotes the  WTe$_2$/SnSe overlap region ($ \approx 40\mu$m$\times 20 \mu$m), which is highly stable even for long-period measurements. We study electron transport across the  SnSe/WTe$_2$ interface (about 200~kOhm resistance value) in a two-point technique by applying voltage $V$ to the contact C6 in respect to the C3 one, and measuring the current $I$ in the circuit. Gate voltage can be applied to the silicon substrate.}
\label{sample}
\end{figure}

WTe$_2$ is usually considered as  a Weyl semimetal~\cite{das16,feng2016}, while now it is also an example of a polar metal with out-of-plane ferroelectric polarization~\cite{WTe}. The polarization is observed even for three-dimensional crystal due to broken inversion symmetry for Td crystal structure~\cite{WTe2_str1,WTe2_str2}. SnSe is a  semiconductor, the  in-plane ferroelectric polarization appears~\cite{Lopez} due to the distortion of centrosymmetric $Pnma$ orthorhombic structure at low thicknesses~\cite{SnSe}. The critical thickness can be estimated~\cite{SnSeprop} as 300~nm for SnSe.

For the present experiment, SnSe compound was synthesized by reaction of selenium vapors with the melt of high-purity tin in evacuated silica ampules.  The SnSe layered single crystal was grown by vertical zone melting in silica crucibles under argon pressure. WTe$_2$ compound was synthesized from elements by reaction of metal with tellurium vapor in the sealed silica ampule. The WTe$_2$ crystals were grown by the two-stage iodine transport~\cite{WTesynth}. Ultra-thin SnSe and WTe$_2$ flakes (about 100-200~nm) are obtained by regular mechanical exfoliation from the initial layered ingot. Thin single-crystal flakes of these materials have been well characterized in transport investigations~\cite{WTeour,SnSeour1,SnSeour2,WTe2shapiro,WTe2chiral}. 

We assemble SnSe/WTe$_2$ heterostructure on the pre-defined Au leads pattern to avoid  chemical or thermal treatment of the initial materials, similarly to single-flake samples~\cite{SnSeour1,SnSeour2,WTe2shapiro,WTe2chiral,black}. 5~$\mu$m separated leads are formed on the standard SiO$_2$ substrate by lift-off technique after thermal evaporation of 100~nm Au.   To obtain separate contacts to the individual layers,  thin SnSe and WTe$_2$ flakes are placed on two independent contact groups (the left and the right ones  in Fig.~\ref{sample}), so Au-SnSe or Au-WTe$_2$ junctions are formed at the bottom surfaces of the individual  flakes. This procedure provides electrically stable contacts with highly transparent metal-semiconductor interfaces, which has been verified for the individual flakes~\cite{SnSeour1,SnSeour2,WTe2shapiro,WTe2chiral,black}. SnSe/WTe$_2$ heterostructure appears as a small (40~$\mu$m$\times 20 \mu$m))  overlap of the flakes at the center of the structure in Fig.~\ref{sample}. Despite the apparent simplicity of the heterostructure fabrication, the samples are stable even for long-period measurements, the observed behavior can be well reproduced for different samples, as it is demonstrated below. Also,  SiO$_2$ substrate protects the flakes from any oxidation/contamination~\cite{black}, so all measurements are taken at room temperature under ambient conditions.

The prepared heterostructure allows to measure electron transport across  SnSe/WTe$_2$ interface. In the case of high resistance samples, one have to use two-point technique with direct application of the voltage bias $V$ to one of the left contacts in Fig.~\ref{sample} in respect to one of the right contacts. We analyze differential conductance $dI/dV$ behavior in dependence on the dc voltage bias. To obtain $dI/dV(V)$ curves, the applied dc bias $V$ is additionally modulated by a small (10 mV) ac component at about 1~kHz frequency. The ac current component is measured by lock-in, being proportional to differential conductance $dI/dV$ at a given dc bias value. We verify that the obtained $dI/dV(V)$ curves are independent of the modulation frequency in the range 1 kHz--10 kHz, which is determined by applied filters. Also, standard oxidized silicon substrate allows to apply the gate voltage $V_g$ to the p-doped silicon across the 100~nm oxide layer. Even for relatively thick (100-200~nm) flakes, ferroelectric polarization is sensitive to the gate electric field~\cite{WTeour}, since the relevant (bottom) flake surfaces are adjoined to the SiO$_2$ surface.

\section{Experimental results}

\begin{figure}[t]
\center{\includegraphics[width=\columnwidth]{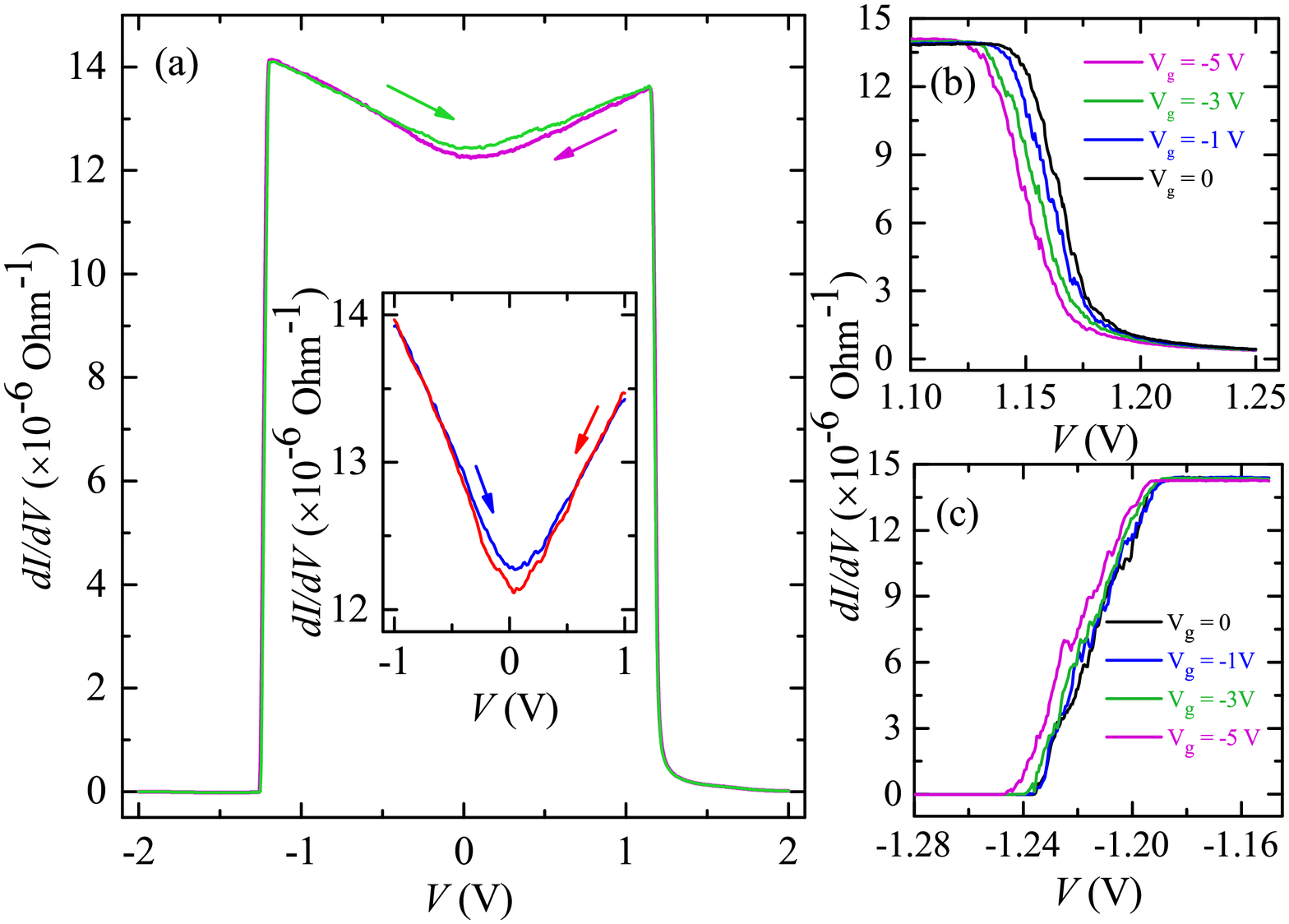}}
\caption{ (Color online) (a) Examples of $dI/dV(V)$ curves for two dc bias $V$ sweep directions. The curves demonstrate abrupt symmetric drop to zero differential conductance at $V_{th}\approx\pm1.2$~V bias values. This $dI/dV$  drop  shows no hysteresis, so $V_{th}$ does not depend on the voltage sweep direction. Inset shows an enlarged low-bias region with small hysteresis around the zero bias, which is known for the individual WTe$_2$ or SnSe flakes~\cite{WTeour,SnSeour1,SnSeour2}. (b,c) Gate voltage  dependence of the $dI/dV(V)$ regions near positive and the negative thresholds. Negative gate voltages decrease the positive $V_{th}$ value and simultaneously increase the negative one. In contrast, the gate electric field does not  noticeably change the $dI/dV$ conductance below the threshold, so the carrier concentration is nearly constant for the applied gate voltages. There is no gate leakage in the present gate voltage range.
}
\label{dI/dV}
\end{figure}

$dI/dV(V)$ behavior is shown in Fig.~\ref{dI/dV} (a) for two dc bias $V$ sweep directions for one of the SnSe/WTe$_2$ heterostructures. At low biases (below $\pm 1.2$~V), $dI/dV(V)$ behavior is nonlinear with small hysteresis around the zero bias. This low-bias region is enlarged in the inset to  Fig.~\ref{dI/dV} (a), 
both the hysteresis and $dI/dV(V)$ behavior are known to originate from ferroelectric polarization for the individual WTe$_2$ and  SnSe flakes~\cite{WTeour,SnSeour1,SnSeour2}.

\begin{figure}[t]
\center{\includegraphics[width=\columnwidth]{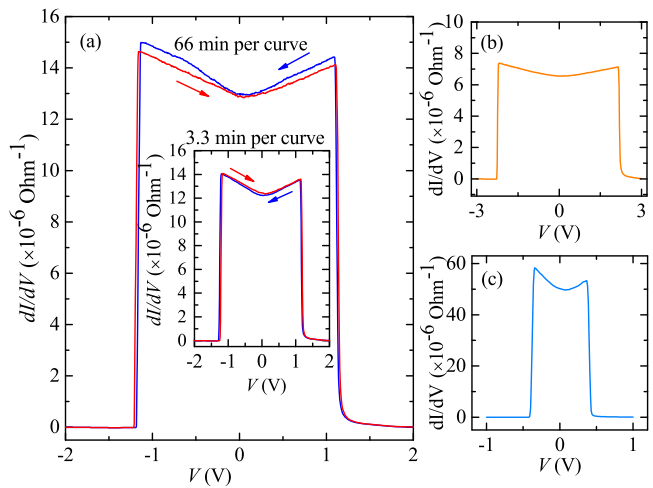}}
\caption{ (Color online) (a) $dI/dV(V)$ curves for different sweep rates (66 minutes per a curve for the main field and 3.3 minutes for the inset) for the same sample as in Fig.~\ref{dI/dV}, where it takes 10 minutes per a curve. The low-bias hysteresis is sensitive to the sweep rate, as it is expected for a single SnSe flake~\cite{SnSeour1,SnSeour2}. However, the threshold values $V_{th}\approx\pm1.2$~V are well stable for different rates. (b,c) Similar $dI/dV(V)$ curves for two other SnSe/WTe$_2$ heterostructures with strongly different initial (zero-bias) conductance values. The curves are symmetric, but $V_{th}$ value is unique for a particular sample.}
\label{diffr samples}
\end{figure}

For the  SnSe/WTe$_2$ heterostructure, our main finding is the  abrupt symmetric $dI/dV$ conductance drop at high biases, which can not be observed for the individual SnSe or WTe$_2$ flakes~\cite{WTeour,SnSeour1,SnSeour2}. In Fig.~\ref{dI/dV} (a), $dI/dV(V)$ curves demonstrate a drop to zero conductance at $V_{th} \approx \pm1.2$~V bias values. This $dI/dV$ conductance drop  shows practically no hysteresis, so $dI/dV$ is abruptly changed at the same threshold voltage value $V_{th}$, irrespectively to the voltage sweep direction. Also, $V_{th}$ value is well reproducible in different voltage sweeps, it is  unique for a particular sample: $V_{th}$ value is inversely proportional to the zero-bias conductance (see below the description of Fig.~\ref{diffr samples}). 

The threshold position can be affected by the gate voltage, see Figs.~\ref{dI/dV} (b) and (c).  Negative gate voltages decrease the  $V_{th}$ value for positive biases and simultaneously increase it for the negative one, so the $dI/dV(V)$ curve is shifted monotonously to negative biases. The effect is well-noticeable, it is about 4\% of the $V_{th}$ value. In contrast, the gate electric field does not change the $dI/dV$ conductance below the threshold, so the carrier concentration is constant for the applied gate voltages. We check, that there is no gate leakage in the present gate voltage range.

\begin{figure}[t]
\center{\includegraphics[width=\columnwidth]{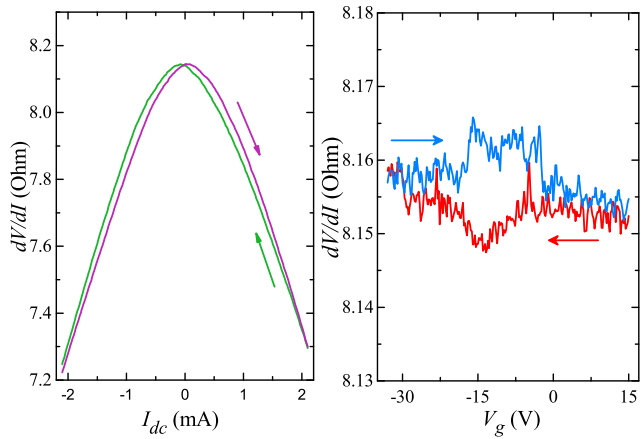}}
\caption{  (Color online) (a) $dV/dI(I)$ curves for the individual WTe$_2$ layer. Source-drain field variation leads to the additional polarization current in WTe$_2$ conductive ferroelectric~\cite{WTe}, which appears as sweep direction dependence in $dV/dI(I)$ curves. (b) Hysteresis of the same magnitude in the gate voltage dependence, which is a standard Sawyer-Tower ferroelectric polarization loop~\cite{ferr_book,scheme}, similarly to the  polarization change by high external electric field in Ref.~\cite{WTe}. The loop center is slightly shifted due to the band bending at the gate dielectric interface.}
\label{WTe}
\end{figure}

For a single SnSe or WTe$_2$ flake, the low-bias hysteresis reflects slow relaxation processes due to the additional polarization current in conductive ferroelectrics~\cite{WTeour,SnSeour1,SnSeour2}.   Fig.~\ref{WTe}  shows this behavior for the individual WTe$_2$ layer. Even well-conducting WTe$_2$ single crystals  demonstrate  ferroelectricity at room temperature, which has been shown by direct visualization of ferroelectric domains~\cite{WTe}. In our setup, there are two possible directions of the external electric fields in the WTe$_2$ layer, the result is depicted in Fig.~\ref{WTe}: (a) the source-drain field $E_{sd}=\rho j$, which is connected with the flowing current, $E_{sd}$ is parallel to the WTe$_2$ surface; (b) The gate field, $E_{gate}=V_g/d$, where $d=300$~nm is the SiO$_2$ oxide thickness, $E_g$ is directed  normally to the WTe$_2$ surface. Any variation of  electric fields leads to the additional polarization current.

In the latter case we observe a standard Sawyer-Tower ferroelectric polarization loop~\cite{ferr_book,scheme}, similarly to the  polarization change by high external electric field in Ref.~\cite{WTe}. The loop center is slightly shifted in Fig.~\ref{WTe} (b) due to the band bending at the gate dielectric interface.  In the former case, source-drain field variation leads to the hysteresis of the same magnitude in Fig.~\ref{WTe} (a).  Similar results can be obtained for the individual SnSe layer~\cite{SnSeour1,SnSeour2}.

We also confirm the hysteresis sweep-rate dependence for low biases in  Fig.~\ref{diffr samples} (a), but the threshold values $V_{th}\approx\pm1.2$~V are well stable for different  rates, low hysteresis can be seen around $V_{th}\approx\pm1.2$~V position only  for the highest sweep rate in the inset to  Fig.~\ref{diffr samples} (a). 

Qualitatively, similar $dI/dV(V)$  behavior can be demonstrated for SnSe/WTe$_2$ heterostructures with strongly different initial conductance, see  Fig.~\ref{diffr samples} (b) and (c). The threshold $V_{th}$ positions are also symmetric, while the $V_{th}$ values are different in the figure: $V_{th}$ is inversely proportional to the zero-bias conductance for all three samples in Fig.~\ref{diffr samples}, which implies constant threshold  source-drain electric field, see the discussion below.

\begin{figure}[t]
\center{\includegraphics[width=\columnwidth]{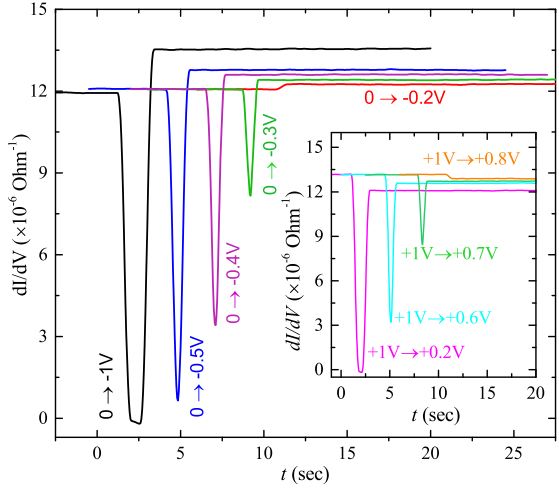}}
\caption{ Re-entrant transition to the low-conductive state for abrupt change of the bias voltage $\Delta V$ for $V<V_{th}$.  For low $\Delta V < 0.2$~V, $dI/dV(t)$ shows monotonous increase in exact correspondence with the $dI/dV(V)$ dependence from  Fig.~\ref{dI/dV} (a), the slow relaxation is insignificant for the present scales. For $\Delta V > 0.3$~V values, this increase goes through the preliminary $dI/dV$ drop.  The drop value grows with  $\Delta V$, so $dI/dV$ goes through the $dI/dV=0$ region for $\Delta V > 0.5$~V. Inset shows similar behavior for another bias $V$ and the sign of the bias change $\Delta V$. Time-dependent $dI/dV(t)$ curves are shifted horizontally for clarity, so the starting point of the bias change coincides with the beginning of the $dI/dV$ drop}
\label{relaxation}
\end{figure}

To our surprise,  SnSe/WTe$_2$ heterostructure shows re-entrant transition to zero-conductance state even at low biases $V<V_{th}$, by abrupt change of the bias value $\Delta V$, as it is shown in Fig.~\ref{relaxation}. Let us start from $V=0$, as depicted in the main field of Fig.~\ref{relaxation}. For low $\Delta V < 0.2$~V, $dI/dV$ is monotonously increased in exact correspondence with the $dI/dV(V)$ curve from  Fig.~\ref{dI/dV} (a), the slow relaxation is insignificant for the scales in Fig.~\ref{relaxation}. For $\Delta V > 0.3$~V, this increase goes through the initial $dI/dV$ drop.  The drop value grows with  $\Delta V$, so differential conductance goes through the $dI/dV=0$ region for $\Delta V > 0.5$~V. This behavior does not depend on the initial bias $V<V_{th}$ or the sign of the bias change $\Delta V$, as it is demonstrated in the inset to Fig.~\ref{relaxation}. Thus, the abrupt change of the bias voltage $V$ leads to re-entrant switching to zero conductance  at low biases $V<V_{th}$, while  the zero-conductance $dI/dV$ state is stable at high biases $V>V_{th}$.

\section{Discussion}

As a result, SnSe/WTe$_2$ heterostructure demonstrates sharp symmetric  drop to zero $dI/dV$ differential conductance at some threshold bias  voltage $V_{th}$, which is sensitive to the gate voltage.  Moreover,  SnSe/WTe$_2$ heterostructure shows re-entrant transition to the low-conductive state for abrupt change of the bias voltage even below the threshold. This behavior is well reproducible for different SnSe/WTe$_2$ samples, while it can not be observed for single SnSe or WTe$_2$ flakes~\cite{WTeour,SnSeour1,SnSeour2}.

First of all, we should exclude sample overheating by the flowing current as the origin of the observed effects. WTe$_2$ crystal structure (Td) is known to be stable in a wide temperature range at ambient pressure~\cite{WTe2_str1,WTe2_str2}. The martensitic phase transition is known~\cite{structSnSe} for SnSe at $\approx$480-530$^\circ$~C, but it is necessarily accompanies by the prominent hysteresis of the transition point~\cite{zhao,SnSeour2}, which is just opposite to the $dI/dV(V)$ behavior in Figs.~\ref{dI/dV} and~\ref{diffr samples}. Also, temperature-induced phase transition is inconsistent with the re-entrant transition to the low-conductive state  in Fig.~\ref{relaxation}. 

Also, any band bending/reconstruction effects at the interface should be strongly asymmetric in respect to the bias sign, while experimental $dI/dV(V)$ curves  are highly symmetric in Figs.~\ref{dI/dV} and~\ref{diffr samples}, so any possible influence of the band bending/reconstruction effects is within the  small difference between the positive and negative thresholds in Fig.~\ref{dI/dV} (b) and (c). It can be estimated as about 0.05 V, the gate voltage effect is of the same value.  We can not expect significant band bending for the well-conducting WTe2 and SnSe materials.

\begin{figure}[t]
\center{\includegraphics[width=\columnwidth]{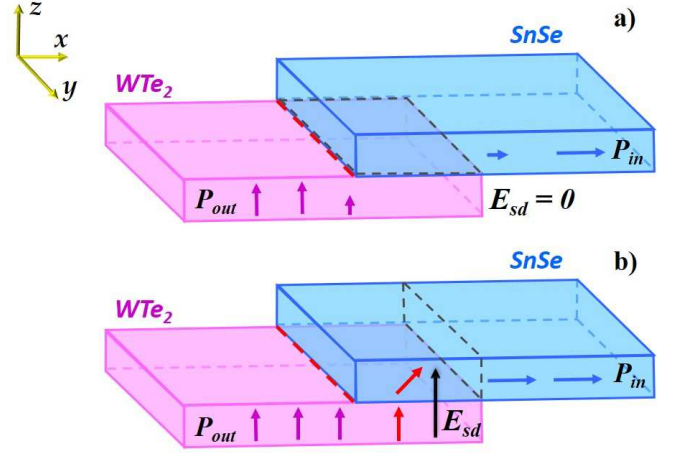}}
\caption{ Schematic diagram for the ferroelectric polarization at the SnSe/WTe$_2$ interface. WTe$_2$ is characterized by out-of-plane ferroelectric polarization~\cite{WTe} $P_{out}$, while it is in-plane oriented ($P_{in}$) in SnSe thin flakes~\cite{SnSeprop}. (a) At zero bias, polarization should be continuously rotated within the overlap region, depicted by the black dashed rectangular. However, WTe$_2$ is characterized by high in-plane conductance~\cite{das16,feng2016}, which should efficiently screen the in-plane electric field component both in WTe$_2$ and SnSe layers due to the proximity effect. Thus, for the macroscopic-size SnSe/WTe$_2$ overlap region, one should expect zero (or strongly diminished) ferroelectric polarizations both in  WTe$_2$ and SnSe layers. (b) At high biases, current-induced electric field $E_{sd}$ enhances the out-of-plane polarization in the WTe$_2$ layer, and, due to the interface coupling~\cite{PM}, it should secondly switch the polarization out-of-plane also in the adjacent SnSe region at some threshold $E_{sd}$ field. The planar contact (about 40~$\mu$m$\times 20 \mu$m) with zero polarization is transformed into the side junction ($\approx 40 \mu$m$\times 100$~nm) with strong polarization mismatch. The latter leads to the significant interface resistance  contribution, which we observe as the strong conductivity drop.}
\label{discussion}
\end{figure}

It is important, that low-bias $dV/dI(I)$ behavior reflects polarization dynamics in  conductive ferroelectirics~\cite{WTeour,SnSeour1,SnSeour2} in Figs.~\ref{dI/dV},~\ref{WTe},  and~\ref{diffr samples}.  Thus, it is quite reasonable to ascribe the observed high-bias behavior to the interface coupling of the ferroelectric polarizations at the SnSe/WTe$_2$ interface, as it was predicted in Ref.~\cite{PM}. WTe$_2$ is characterized by semimetal spectrum~\cite{das16,feng2016} with out-of-plane ferroelectric polarization~\cite{WTe}, so it is a good representation of the polar metal concept~\cite{PM,pm1,pm2,pm3,pm4}. SnSe thin flake is a ferroelectric semiconductor with in-plane polarization, while the SnSe conductivity is significant at room temperature~\cite{SnSeprop}.  Thus, SnSe/WTe$_2$ bilayer can be considered as a heterostructure between two conducting ferroelectrics with different polarization directions~\cite{PM}, as it is schematically depicted in Fig~\ref{discussion} (a).

Due to the interface coupling, one can expect complicated polarization structure at the SnSe/WTe$_2$ interface:  polarization should be continuously rotated within the overlap region~\cite{PM}. However, WTe$_2$ is characterized by high in-plane conductance~\cite{das16,feng2016}, which should efficiently screen the in-plane electric field component both in WTe$_2$ and SnSe layers due to the proximity effect.  Thus, for the macroscopic-size SnSe/WTe$_2$ overlap region (depicted by black dashed rectangular in Fig~\ref{discussion} (a), area is about 40~$\mu$m$\times 20 \mu$m), one should expect zero (or strongly diminished) ferroelectric polarizations both in  WTe$_2$ and SnSe layers, as it is shown in Fig~\ref{discussion} (a),  so the polarization mismatch has low (or even zero) influence on the sample resistance. This conclusion is also supported by low-bias measurements. In our samples, the total resistance (e.g., between C3 and C6 in Fig.~\ref{sample})  consists of the in-series connected resistances of the SnSe/WTe$_2$ interface and the resistances of the SnSe and WTe$_2$ layers. The individual WTe$_2$ flake is of 10 -- 100 Ohm resistance~\cite{WTeour} (e.g. as measured between C3 and C4 contacts), the single SnSe layer is characterized~ by  50 -- 200 kOhm values between C5 and C6, so there is no observable interface contribution. Low variation of  the source-drain bias leads to the additional polarization current in the bulk SnSe ($\rho_{SnSe}>>\rho_{WTe_2}$), which we observe~\cite{WTeour,SnSeour1,SnSeour2} as low-bias hysteresis in Fig.~\ref{dI/dV}. 

To understand the high-bias switching, it is important that current-induced electric field $E_{sd}$ is oriented normally to the SnSe/WTe$_2$ interface due to the high WTe$_2$ in-plane conductance. It enhances the out-of-plane polarization in the WTe$_2$ layer, and, due to the interface coupling~\cite{PM}, it should secondly switch the polarization out-of-plane also in the adjacent SnSe region at some threshold $E_{sd}$ field. The latter can be estimated from the experimental $V_{th}$ values  as  $E_{sd}\sim V_{th} \approx 10^4-10^5$~V/m. In this case, the planar contact (depicted by black dashed rectangular in Fig~\ref{discussion} (a), area is about 40~$\mu$m$\times 20 \mu$m) is transformed into the side junction in Fig~\ref{discussion} (b) with $\approx 40\mu$m$\times 100$~nm area. The polarization mismatch leads to the significant interface resistance  contribution, analogously to the scattering of the charge carriers on the domain walls~\cite{domain wall effect on transp}.

This effect is not sensitive to the bias sign, since both $E_{sd}$ directions at the interface can drive 90$^\circ$ change of the initial SnSe in-plane polarization~\cite{PM}. Similarly to $E_{sd}$, the gate electric field $E_g$ is directed  normally to the SnSe/WTe$_2$ interface. Thus, it increases the interface field for one $E_{sd}$ direction and decreases it for the opposite one, as we observe in Fig.~\ref{dI/dV} (b,c). The proposed model is also confirmed by the excellent  stability of the threshold regions in the  experimental curves, see Figs.~\ref{dI/dV}, and~\ref{diffr samples}. Any contact or scattering effects should demonstrate random fluctuations from sample to sample. In contrast, $V_{th}$ is inversely proportional to the zero-bias conductance for three different samples in Fig.~\ref{diffr samples}, which implies constant, device-independent,  threshold electric field  $E_{sd}$.

The above picture is also confirmed by the time-dependent curves in Fig.~\ref{relaxation}. At constant bias $V$, the WTe$_2$ flake is nearly equipotential because of low resistivity  $\rho_{WTe_2}<<\rho_{SnSe}$. Thus, the abrupt change in the bias $\Delta V$ is applied initially at the SnSe/WTe$_2$ interface, while  it is redistributed afterwards over the resistive SnSe flake. It, therefore, temporary forces the transition to the low-conducting state, which can not be preserved after the field redistribution at $V<V_{th}$. In Fig.~\ref{relaxation}, re-entrant transition indeed depends on the $\Delta V$ value, so the time-dependent behavior confirms our interpretation of the SnSe/WTe$_2$ polarization state control  by electron transport through the interface.

\section{Conclusion}
As a conclusion, we fabricate SnSe/WTe$_2$  heterostructure to  experimentally investigate charge transport between two ferroelectric van der Waals materials with different polarization directions. WTe$_2$ is a polar metal with out-of-plane ferroelectric polarization, while SnSe ferroelectric semiconductor is polarized in-plane, so one should expect complicated polarization structure at the SnSe/WTe$_2$ interface. We study $dI/dV(V)$ curves, which  demonstrate sharp symmetric  drop to zero $dI/dV$ differential conductance at some threshold bias  voltages $\pm V_{th}$, which are nearly symmetric in respect to the bias sign. While the gate electric field is too small to noticeably affect the carrier concentration, the positive and negative threshold positions are sensitive to the gate voltage. Also,  SnSe/WTe$_2$ heterostructure shows re-entrant transition to the low-conductive $dI/dV=0$ state for abrupt change of the bias voltage even below the threshold values. This behavior can not be observed for single SnSe or WTe$_2$ flakes, so we interpret it as  a result of the SnSe/WTe$_2$ interface coupling. In this case, some threshold value of the   electric field at the SnSe/WTe$_2$ interface is  enough to  drive 90$^\circ$ change of the initial SnSe in-plane polarization in the overlap region. The polarization mismatch leads to the significant interface resistance  contribution, analogously to the scattering of the charge carriers on the domain walls.   Thus, we demonstrate  polarization state control  by electron transport through the SnSe/WTe$_2$ interface. 

\section{Acknowledgement}

We wish to thank S.S~Khasanov for X-ray sample characterization.  We gratefully acknowledge financial support  by the  Russian Science Foundation, project RSF-22-22-00229, https://rscf.ru/project/22-22-00229/.

\end{document}